
\hfuzz=2pt

\def\tr{{\rm tr}}

\font\titlefont=cmbx10 scaled\magstep1
\magnification=\magstep1

\null
\rightline{SISSA 59/95/EP}
\rightline{\tt hep-th/9505172}
\vskip 1.5cm
\centerline{\titlefont THE HEAT-KERNEL}
\smallskip
\centerline{\titlefont AND THE AVERAGE EFFECTIVE POTENTIAL}
\vskip 1.5cm
\centerline{\bf R. Floreanini \footnote{$^*$}{\tt florean@ts.infn.it}}
\smallskip
\centerline{Istituto Nazionale di Fisica Nucleare, Sezione di
Trieste}
\centerline{Dipartimento di Fisica Teorica, Universit\`a di Trieste}
\centerline{Strada Costiera 11, 34014 Trieste, Italy}
\bigskip\smallskip
\centerline{\bf R. Percacci \footnote{$^{**}$}{\tt
percacci@tsmi19.sissa.it}}
\smallskip
\centerline{International School for Advanced Studies, Trieste,
Italy}
\centerline{via Beirut 4, 34014 Trieste, Italy}
\centerline{and}
\centerline{Istituto Nazionale di Fisica Nucleare,
Sezione di Trieste}
\vskip 1.8cm
\centerline{\bf Abstract}
\smallskip
\midinsert
\narrower\narrower\noindent
We discuss the definition of the average effective action in terms
of the heat-kernel. As an example we examine a model describing a
self-interacting scalar field, both in flat and curved background,
and study the renormalization group flow of some of the parameters
characterizing its effective potential. Some implications of the
running of these parameters for inflationary cosmology are also
briefly discussed.
\endinsert
\bigskip
\vskip 1cm
\vfil\eject

{\bf 1.} The average effective action $\Gamma_k$ is the quantum action
that describes accurately the behaviour of a given physical system
at the momentum scale $k$. It is obtained via the path integral by
integrating in subsequent steps the fields with momenta larger than $k$.
The scale $k$ effectively acts as an infrared cutoff and the flow of
$\Gamma_k$ as $k$ varies is the renormalization group flow.

Originally, these ideas were applied to spin models on the lattice, [1]
but they have been subsequently adapted to the continuum, where they
have been used to clarify and simplify the notion of renormalizability
in quantum field theory. [2, 3] From the computational point of view, the
implementation of these ideas presented in [4] is very effective.
It has been applied to the study of
the phase structures of lower dimensional models
as well as to simplified versions of QCD, [5] and to quantum gravity. [6-8]

The purpose of this letter is to elaborate on the definition of the effective
action $\Gamma_k$ using heat-kernel techniques. Within this framework,
we shall study the renormalization group flow of some basic parameters
characterizing $\Gamma_k$ in the case of a self-interacting scalar field
theory, both in flat and curved spacetime. The analysis of this model is of
relevance also for the cosmic evolution of the universe. Indeed, the running
of the corresponding average effective potential can have some effects
in inflationary cosmology. A brief discussion on some of these effects
is reported at the end.

\bigskip

{\bf 2.} Let us consider the theory describing
a scalar field $\phi(x)$ interacting
via a potential $V(\phi)$. The usual Euclidean effective
action $\Gamma$ is obtained
from the functional integral by means of a Legendre transform, using
the background field method. At one-loop level, this amounts to the
computation of a determinant,
$$\Gamma(\varphi)={1\over2}\ln\det{\cal O}(\varphi)\ ,\eqno(1)$$
where ${\cal O}(\varphi)$ is the elliptic operator that describes the small
fluctuations of the quantum field $\phi$ around the background configuration
$\varphi$.

The definition of the determinant in (1) requires a regularization, due to
the presence of ultraviolet infinities. We assume that some cutoff $\Lambda$
has been introduced to make (1) meaningful. As we shall see, the choice
of regularization will play no role in the considerations that follow.

In general, the average effective action $\Gamma_k$ can be obtained from
$\Gamma$ by introducing an infrared cutoff at the momentum scale $k$.
The use of a sharp cutoff has some disadvantages; as described in Ref.[4],
a smooth cutoff is preferable. To this end, introduce the function
$$P_k(q^2)={q^2\over 1-e^{-2a(q^2/k^2)^b}}\ ,\eqno(2)$$
with $a$ and $b$ positive constant parameters.
For $|q|\gg k$, the function $P_k(q^2)$
exponentially approaches $q^2$; on the other hand, it tends to $k^2$
(when $b=1$), or diverges (when $b>1$) for $q^2\sim0$.

Consider now the perturbative expansion of $\Gamma$ in (1)
in terms of one-loop graphs; each graph in this expansion involves an
integration over a momentum $q$. The average effective action $\Gamma_k$
is then obtained from (1) by making in each momentum integral
the substitution:
$$ q^\mu\rightarrow q^\mu\sqrt{P_k(q^2)\over q^2}\ .\eqno(3)$$
In this way the propagation of the modes with momenta smaller than the
scale $k$ is suppressed. Note that for $b\rightarrow\infty$, the function
$e^{-2a(q^2/k^2)^b}$ approaches a step function; in this case the modes
with $|q|<k$ do not propagate: we are in the case of a sharp infrared
cutoff. In conclusion, one can write
$$\Gamma_k(\varphi)={1\over2}\ln {\det}_k{\cal O}(\varphi)\ ,\eqno(4)$$
where $\det_k$ is the determinant with its one-loop momentum integrations
modified as in (3).

A more general definition for $\Gamma_k$ can be given. It is based
on the heat-kernel or proper-time definition of the determinant in (1). [9]
To this end, let us introduce the heat equation for the operator $\cal O$:
$${d\over dt}K+{\cal O}\cdot K=\ 0\ ,\qquad K(x,y)|_{t=0}=\delta(x,y)\ ;
\eqno(5)$$
it has the formal solution: $K=e^{-t{\cal O}}$. Then, one can write:
$$\Gamma=-{1\over2}\int_{1/\Lambda^2}^\infty {dt\over t}\,
\tr K\ ,\eqno(6)$$
with $\Lambda$ the ultraviolet cutoff. As in the perturbative definition
of the average effective action given in (4), also in the present
case one passes from (6) to $\Gamma_k$ by introducing an infrared cutoff $k$.
This can be effectively achieved by multiplying the integrand in the rhs of
(6) by a universal $k$-dependent function $F_k(t)$:
$$\Gamma_k=-{1\over2}\int_{1/\Lambda^2}^\infty {dt\over t}\, F_k(t)
\, \tr K\ .\eqno(7)$$

The function $F_k$ has obviously to satisfy some general conditions.
First of all, note that dimensional arguments require $F_k(t)$  to depend
only on the dimensionless variable $z=k^2 t$: $F_k(t)=F(k^2 t)$.
Further, since $\Gamma_k=\Gamma$ at $k=\,0$, one must have $F(0)=1$.
Also, $F(z)$ must tend to zero sufficiently rapidly for large $z$ to suppress
the small momentum modes. Actually, to enforce this requirement one needs
a condition also on the first derivative of $F(z)$:
$F^\prime(z)= -z^2\, f(z)$, with $f(z)$ a positive regular function in the
neighborhood of the origin. This last condition is technical, and
assures that the renormalization group equation for $\Gamma_k$ be
ultraviolet finite. This is an obvious requirement, since the
ultraviolet divergences in the effective action should not be
affected by the presence of the cutoff in the infrared region
(see also [4]).

\vfill\eject

{\bf 3.} As a first example of application of the definition (7),
let us study the renormalization group flow
of the effective potential for a self-interacting
scalar field in flat space.
Assuming a classical potential
$V(\varphi)=\lambda\big(\varphi^2-\varphi^2_m\big)^2/2$, the small
fluctuation operator takes the form:
$${\cal O}(\varphi)=-\partial^2+2\lambda(3\varphi^2-\varphi_m^2)\ .\eqno(8)$$
The running potential is also assumed to have the form
$$V_k(\varphi)={1\over2}\lambda_k\big(\varphi^2-\varphi_k^2\big)^2\ ;
\eqno(9)$$
``irrelevant'' terms will be neglected.

We want now to study the evolution in $k$ of the minimum $\varphi_k$
and of the coupling constant $\lambda_k$, which are defined by
$V_k'(\varphi_k)=\, 0$ and $\lambda_k=V_k''(\varphi_k)$ (a prime
signifies derivative with respect to $\varphi^2$).
This is determined by the following coupled differential equations:
$$\eqalign{&k{\partial\varphi_k^2\over\partial k}=k^2\, \alpha(k)\ ,
\phantom{\bigg|_{\varphi_k}}\cr
         &k{\partial\lambda_k\over\partial k}=\beta(k)\ ,
\phantom{\bigg|_{\varphi_k}}}\qquad
  \eqalign{&\alpha(k)=-{1\over k^2\lambda_k}\,
k{\partial V_k^\prime\over\partial k}\bigg|_{\varphi_k}\ ,\cr
         &\beta(k)=k{\partial V_k''\over\partial k}\bigg|_{\varphi_k}\ .}
\eqno(10)$$
Let us stress that these are approximated evolution equations, consistent
with the choice (9). For example a triple derivative $V_k'''$,
taking into account the $k$ dependence of the point of definition of
$\lambda_k$, has been dropped from $\beta(k)$.

Now, from the definition (7) and recalling that for constant
background $\varphi$, $\Gamma_k(\varphi)=\int d^4x\, V_k(\varphi)$, one obtains
$$\eqalign{V_k(\varphi)=&-{1\over2}\int_{1/\Lambda^2}^\infty {dt\over t}\,
F(k^2 t)\, \int {d^4q\over(2\pi)^4}\,
e^{-t\big[q^2+2\lambda(3\varphi^2-\varphi^2_m)\big]}\ ,\cr
                       =&-{1\over 32\pi^2}\int_{1/\Lambda^2}^\infty
{dt\over t^3}\, F(k^2 t)\, e^{-2t\lambda(3\varphi^2-\varphi^2_m)}\ .}
\eqno(11)$$
{}From this expression one can compute the functions $\alpha(k)$ and $\beta(k)$
by taking the appropriate derivatives, and by substituting for the
parameters entering the classical potential, $\lambda$, $\varphi_m$, and
for the background $\varphi$, their running counterparts $\lambda_k$
and $\varphi_k$. This is the renormalization group improvement, that amounts
to recursively upgrade the effective potential at any scale $k$
(for further details, see [4, 5]). Explicitly, one finds:
$$\eqalignno{&\alpha(k)={3\over8\pi^2}\, \lambda_k\int_0^\infty
{dz\over z}\, e^{-4z\lambda_k(\varphi_k^2/k^2)}\, F^\prime(z)\ , &(12a)\cr
             &\beta(k)=-{9\over4\pi^2}\, \lambda_k^2\int_0^\infty dz\,
e^{-4z\lambda_k(\varphi_k^2/k^2)}\, F^\prime(z)\ . &(12b)}$$

Actually, we are interested in approximated expressions for $\alpha$
and $\beta$, valid in certain regimes, so that analytic solutions for the
system (10) of flow equations can be explicitly obtained. First, let us
assume that $\varphi_k^2/k^2\ll 1$. In this case, one finds:
$$\eqalign{&k{\partial\varphi_k^2\over\partial k}= {3A\over8\pi^2}\, k^2\ ,\cr
           &k{\partial\lambda_k\over\partial k}={9\over4\pi^2}\,
\lambda_k^2\ ,}\eqno(13)$$
where: $A=\int_0^\infty dz\, z\, f(z)$. Note that the dependence on the
cutoff function $F(z)$ is only via the constant $A$;
in particular, the value of $\beta$
is universal. This is to be expected since in this regime one should recover
the results of the standard perturbation theory. Indeed, the equations
(13) gives a logarithmic running for quartic self-coupling, while the
position of the minimum $\varphi_k$ scales with $k$ as its dimension,
up to logarithmic corrections.

In the region for which $\varphi_k^2/k^2\gg 1$, one obtains:
$$\eqalign{&k{\partial\varphi_k^2\over\partial k}= {3B\over128\pi^2}\,
{k^2\over\lambda_k^2}\, \left({k^2\over\varphi_k^2}\right)^2\ ,\cr
           &k{\partial\lambda_k\over\partial k}={9B\over128\pi^2}\,
{1\over\lambda_k}\, \left({k^2\over\varphi_k^2}\right)^3\ ,}\eqno(14)$$
with $B=f(z)|_{z=0}$. In this regime, $\varphi_k$ and $\lambda_k$
tend to constants as $k$ goes to zero. This behaviour and
the one dictated by (13) coincide with the ones computed in Refs.[4, 8],
using standard methods.

The renormalization flow equations (13) and (14) depend on the choice
of the infrared cutoff function $F$ through the constants $A$ and $B$.
A particularly interesting explicit choice for $F$ can be obtained
using the definition (2), (3), (4) for the average effective action.
In this case one obtains:
$$F(z)=\int_0^\infty dx\, x\, \exp\left[-{x\over1-e^{-2a(x/z)^b}}\right]
\ .\eqno(15)$$
This expression essentially coincides with the trace of the heat kernel
for the operator $P_k(-\partial^2)$.
One can check that this function satisfies all the properties we have
previously required for $F$. With this particular choice, one
easily gets:
$$ A={1\over(2a)^{1/b}}\, \Gamma(1+1/b)\ ,\qquad
   B={1\over(2a)^{3/b}}\, \Gamma(1+3/b)\, \zeta(3/b)\ .\eqno(16)$$

\bigskip

{\bf 4.} The definition (7) for the average effective action
has the advantage that it can be equally
well applied to field theory models on curved spacetime backgrounds.
As an example, let us consider the theory for a scalar
field $\varphi$, with Euclidean action:
$$S(\varphi)=\int d^4x \sqrt{g}\left({1\over2}g^{\mu\nu}
\partial_\mu\varphi\, \partial_\nu\varphi+{1\over2} \xi\, R\,\varphi^2+
V(\varphi)\right)\ ,\eqno(17)$$
propagating on a background with metric $g_{\mu\nu}$ and scalar curvature
$R$. We take again a potential $V$ of the form (9) and concentrate
on the study of the renormalization group flow of the corresponding
parameters $\lambda_k$, $\xi_k$ and of the minimum $\varphi_k$.

The explicit expression for $\Gamma_k$ can be obtained applying Eq.(7),
where now $K$ is the heat kernel for the operator
$${\cal O}(\varphi)=\widetilde{\cal O}+2\lambda(3\varphi^2-\varphi_m^2)\ ,
\qquad \widetilde{\cal O}=-\nabla^2+\xi\, R\ .\eqno(18)$$
More explicitly, calling $\widetilde K$ the heat kernel for
$\widetilde{\cal O}$, one has
$$\Gamma_k(\varphi)=-{1\over2}\int_{1/\Lambda^2}^\infty {dt\over t}\,
F(k^2 t)\, e^{-2t\lambda(3\varphi^2-\varphi_m^2)}\, \tr\widetilde K\ .
\eqno(19)$$
In this case, the trace of $\widetilde K$ can not be computed in closed
form. However, it has a Laurent series expansion around $t=\,0$:
$$\tr\widetilde K=\sum_{n=0}^\infty t^{n-3}\int d^4x\, \sqrt{g}\ \tilde b_n\ ,
\eqno(20)$$
where $\tilde b_n$ are the so called heat kernel coefficients. [9]
They can be explicitly
computed in terms of powers of the curvature tensors and their derivatives,
{\it e.g.} $\tilde b_0=1/16\pi^2$, $\tilde b_1=(1/16\pi^2)(1/6-\xi)\, R$.

Notice that in writing (19), we have implicitly taken $F$, the function that
introduces the infrared scale $k$, to be independent of the background metric
$g_{\mu\nu}$. Although other approaches are certainly conceivable,
this choice allows an easy evaluation of the various
beta-functions.

We shall concentrate on the study of the effective potential $V_k(\varphi)$
which is obtained by taking a constant background configuration $\varphi$.
Also, for the sake of simplicity,
we shall take the curvature $R$ to be constant.
In this situation, the parameters $\varphi_k$ and $\lambda_k$ are defined
as before, while we introduce the non-minimal coupling $\xi_k$ as the
derivative with respect to $R$ of $V_k^\prime(\varphi_k)$, at $R=0$.
Then, the momentum flows of $\varphi_k$ and $\lambda_k$ are given again by
the equations (10), while for $\xi_k$ one finds:
$$k{\partial\xi_k\over \partial k}=\gamma(k)\ ,\qquad
\gamma(k)=k{\partial\over\partial k}
{\partial V_k^\prime\over\partial R}\bigg|_{\varphi_k,\ R=0}\ .\eqno(21)$$
The expressions for the quantities $\alpha(k)$, $\beta(k)$ and $\gamma(k)$
can now be explicitly computed as a series expansion in $R$. One finds:
$$\eqalignno{&\alpha(k)=6\lambda_k\sum_{n=0}^\infty
\, {b_n(\xi_k)\over (k^2)^n}\int_0^\infty dz\, z^{n-1}\,
e^{-4z\lambda_k(\varphi_k^2/k^2)}\, F^\prime(z)\ , &(22a)\cr
             &\beta(k)=-36\lambda_k^2\sum_{n=0}^\infty
\, {b_n(\xi_k)\over (k^2)^n}\int_0^\infty dz\, z^n\,
e^{-4z\lambda_k(\varphi_k^2/k^2)}\, F^\prime(z)\ , &(22b)\cr
             &\gamma(k)={3\lambda_k\over4\pi^2}\left({1\over6}-\xi_k\right)
\int_0^\infty dz\,e^{-4z\lambda_k(\varphi_k^2/k^2)}\, F^\prime(z)\ .
&(22c)}$$

As in the case of flat space, the flow equations (10) and (21) simplify
if one takes the limit for which $\varphi_k^2/k^2$ is very small
or very large, and further assume that the curvature $R$ to be small with
respect to $\varphi_m$ and $k$.
Let us first consider the case $\varphi_k^2/k^2\ll 1$.
Keeping only the dominant terms and
defining the convenient
combination $\eta_k=1/6-\xi_k$, one can write:
$$\eqalignno{&k{\partial\varphi_k^2\over\partial k}=
{3\over8\pi^2}\, k^2 \left[A+\eta_k {R\over k^2}\right]\ , &(23a)\cr
             &k{\partial\lambda_k\over\partial k}=
{9\over4\pi^2}\, \lambda_k^2\, \left[1+C\eta_k\, {R\over k^2}\right]
\ , &(23b)\cr
             &k{\partial\eta_k\over\partial k}=
{3\over8\pi^2}\, \lambda_k\eta_k\ , &(23c)}$$
where $C=-\int_0^\infty dz\, F(z)$. Notice that for $R=\,0$ the equations
$(23a)$ and $(23b)$ reduce to those in (13). Further, the value of
$\gamma$, {\it i.e.} the rhs of $(23c)$, reproduces the standard perturbative
result, [10] as expected.
The equations (23) give the dominant logarithmic running
for the coupling constants $\lambda_k$ and $\xi_k$, plus subleading
logarithmic corrections depending on the curvature. Similarly, $\varphi_k^2$
scales as $k^2$, with additional logarithmic corrections proportional
to $R$.

In the complementary region, $\varphi_k^2/k^2\gg 1$, one obtains instead
$$\eqalignno{&k{\partial\varphi_k^2\over\partial k}=
{3B\over128\pi^2}\, {k^2\over\lambda_k^2}\,
\left({k^2\over\varphi_k^2}\right)^2\, \left[1+{\eta_k\over2\lambda_k}\,
{R\over\varphi_k^2}\right]\ , &(24a)\cr
             &k{\partial\lambda_k\over\partial k}=
{9B\over128\pi^2}\, {1\over\lambda_k}\,
\left({k^2\over\varphi_k^2}\right)^3\, \left[1+{3\eta_k\over4\lambda_k}\,
{R\over\varphi_k^2}\right]\ , &(24b)\cr
             &k{\partial\eta_k\over\partial k}=
{3B\over256\pi^2}\, {1\over\lambda_k^2}\,
\left({k^2\over\varphi_k^2}\right)^3\, \eta_k\ . &(24c)}$$
The running of $\varphi_k^2$, $\lambda_k$ and $\xi_k$ is suppressed by
powers of $k^2/\varphi_k^2$, and actually stops for $k\sim 0$. For small $k$,
one can approximate the couplings $\lambda_k$ and $\xi_k$ with their values
at $k=\,0$. In this case, one further has:
$$\varphi_k^2=\varphi_0^2\left\{1+{B\over 256\pi^2}\, {1\over\lambda_0^2}\,
\left({k^2\over\varphi_0^2}\right)^3\left[1+{\eta_0\lambda_0\over2}
\left({R\over\varphi_0^2}\right)\right]\right\}\ .\eqno(25)$$

\bigskip

{\bf 5.} The previous results on the running of the minimum $\varphi_k$
of the effective potential could have some consequences on the cosmic
evolution of the universe. In standard inflationary cosmology, the
field $\varphi$ is a Higgs field of a Grand Unification Model,
that undergoes a spontaneous symmetry breaking. It is responsible
of inflation via its slow rolling from the configuration $\varphi=\, 0$
to the true minimum of the effective potential.

Usually, one treats this evolution semiclassically, by adding the Einstein
term to the action (17), and assuming for the metric $g_{\mu\nu}$ a
Friedman form. The approximated cosmic-time evolution of the scale factor
$a(t)$ of the universe is then given by $(H\equiv\dot a/a)$: [11]
$$H^2={8\pi G_N\over 3}\, V(0)\ ,\eqno(26)$$
where $G_N$ is Newton's constant and $V(0)$ is the value of the effective
potential for the matter field $\varphi$ at the starting point $\varphi=\, 0$.
We use the convention that gives to $a$ the dimensions of length.

In the standard approach, $V(0)$ is considered constant, so that from (26)
one has the inflationary exponential behaviour:
$$ a(t)\sim a(0)\, e^{H\, t}\ .\eqno(27)$$

However, things change if we take into account the running of the effective
potential. In this case, from (9) one has $V_k(0)=\lambda_k\varphi_k^4/2$.
As previously explained, $k$ must coincide with the characteristic scale
of the phenomena under study. We are following the evolution of the
scale factor $a$ of the universe; it is therefore natural to identify:
$k\sim 1/a$. Further, when inflation starts, $k$ is of the order
of Planck's mass $M_P$, and therefore very large with respect to the masses
of all ordinary particles. Indeed, we are studying the cosmic evolution
immediately after Planck's time, {\it i.e.} in the regime where semiclassical
considerations are justified. As shown previously, for such large values
of $k$, $\varphi_k$ scales approximately as $k$,
with a proportionality constant $c$ of order one,
while $\lambda_k$ runs logarithmically.
Neglecting for simplicity this mild change of $\lambda_k$, Eq.(26)
becomes:
$$ H^2={4\pi\lambda G_N\over3}\, {c^4\over a^4}\ .\eqno(28)$$
In this case, one no longer has an inflationary behaviour, since:
$a(t)\sim t^{1/2}$.

This conclusion, based on the simplified equation (26), is a little crude,
but clearly indicates a change in the evolution of the cosmic factor $a(t)$
due to the running of the effective potential for $\varphi$.
More accurate conclusions can be obtained by considering
together the equations of motion of both the field $\varphi$ and the scale
factor $a$. As a starting action we take again (17) plus the Einstein term,
but for simplicity we limit our considerations to the case $\xi=\, 0$.

Assuming an inflationary regime, for which $|\dot H|\ll H^2$, and a slow
changing homogeneous field $\varphi$, one has: [11]
$$\eqalignno{ &3 H\dot\varphi + 2\lambda\varphi\big(\varphi^2-\varphi_k^2\big)
=\, 0\ , &(29a)\cr
            &H^2={4\pi\lambda G_N\over3}\, \big(\varphi^2-\varphi_k^2\big)^2
\ . &(29b)}$$
Here again we neglect the running of the coupling constant $\lambda$.
Irrespectively of the value of $\varphi_k$, the evolution of $\varphi$
is exponentially dumped:
$$\varphi(t)=\varphi(0)\, e^{-h\, t}\ ,\qquad
h=\sqrt{\lambda\over3\pi\, G_N}\ .\eqno(30)$$
Then, using again $\varphi_k=c/a$, $(29b)$ gives
the following behaviour for $a(t)$:
$$\eqalign{a^2(t)= & a^2(0)\,
e^{2\pi G_N \left(\varphi^2(0)-\varphi^2(t)\right)}\cr
&\times\left\{1-{2\pi c^2 G_N\over a^2(0)} e^{-2\pi G_N\varphi^2(0)}\, \left[
E\Big(2\pi G_N\varphi^2(0)\Big)-E\Big(2\pi G_N\varphi^2(t)\Big)
\right]\right\}\ ,}\eqno(31)$$
where $E(x)$ is the exponential-integral: $E(x)=\int^x dz\, e^z/z$.

At the starting point, immediately after Planck's time, one can reasonably
assume: $V_k(\varphi(0))\sim M_P^4$. For $\lambda$ small, this implies that
$\varphi(t)$ starts at a value $\varphi(0)\sim \lambda^{-1/2}\, M_P \gg M_P$,
and varies very little over the interval:
$\Delta t\sim (\sqrt\lambda\, M_P)^{-1}$. In this interval, the scale
function $a(t)$ has an inflationary exponential behaviour due to the first
factor in (31), which is however stopped immediately after by the second
factor.

In this framework there is no need for an additional mechanism to
stop inflation: everything is already encoded in the approximated
equations (31), provided the renormalization flow of $\varphi_k$
is taken into account.
Further detailed analysis is certainly needed to confirm this behaviour.
Nevertheless, we find these preliminary results worth of consideration.

\vskip 1cm

\centerline{\bf References}
\medskip
\noindent
\item{1.} K.G. Wilson and J.B. Kogut, Phys. Rep. {\bf 12C}
(1974) 75;\hfil\break
K.G. Wilson, Rev. Mod. Phys. {\bf 47} (1975) 774
\smallskip
\item{2.} J. Polchinski, Nucl. Phys. {\bf B 231} (1984) 269
\smallskip
\item{3.} B. Warr, Ann. of Phys (NY) {\bf 183} (1988) 1 \hfil\break
C. Becchi, ``On the construction of renormalized quantum field theory
using renormalization group techniques'', in {\it Elementary
Particles, Field Theory and Statistical Mechanics},
M. Bonini, G. Marchesini and E. Onofri, eds.,
University of Parma (1993);\hfil\break
M. Bonini, M. D'Attanasio and G. Marchesini, Nucl. Phys. {\bf B 418}
(1994) 81; {\it ibid.} {\bf B 421} (1994) 429;
{\it ibid.} {\bf B 437} (1995) 163
\smallskip
\item{4.} C. Wetterich, Nucl. Phys. {\bf B 334} (1990) 506;
{\it ibid.} {\bf B 352} (1991) 529; Phys. Lett. {\bf B301} (1993) 90
\smallskip
\item{5.} N. Tetradis and C. Wetterich, Nucl. Phys. {\bf B 398} (1993) 659;
Int. J. Mod. Phys. {\bf A9} (1994) 4029;\hfill\break
M. Reuter and C. Wetterich, Nucl. Phys. {\bf B 391} (1993) 147;
{\it ibid.} {\bf B 417} (1994) 181; {\it ibid.} {\bf B 427} (1994) 291;
\hfill\break  C. Wetterich, Nucl. Phys. {\bf B 423} (1994) 137
\smallskip
\item{6.} R. Floreanini and R. Percacci, Nucl. Phys. {\bf B 436} (1995) 141
\smallskip
\item{7.} R. Floreanini and R. Percacci, The renormalization group flow
of the dilaton potential, Phys. Rev. D, to appear
\smallskip
\item{8.} L. Griguolo and R. Percacci, The beta functions of a scalar theory
coupled to gravity, SISSA-preprint, 1995
\smallskip
\item{9.} B.S. De Witt, Phys. Rep. {\bf 19} (1975) 295;\hfill\break
S.W. Hawking, Comm. Math. Phys. {\bf 55} (1977) 133;\hfill\break
N.D. Birrel and P.C.W. Davies, {\it Quantum Fields in Curved Space},
(Cambridge University Press, Cambridge, 1982)
\smallskip
\item{10.} I.L. Buchbinder, S.D. Odintsov and I.L. Shapiro, {\it Effective
Action in Quantum Gravity}, (IOP, Bristol(UK), 1992)
\smallskip
\item{11.} A. Linde, {\it Particle Physics and Inflationary Cosmology},
(Harwood, 1990)

\bye